\documentclass[preprint,12pt,a4paper]{elsarticle}

\usepackage{amssymb}

\journal{J. Mag. Mag. Mater.}

\begin{document}

\begin{frontmatter}

\title{New room-temperature ferromagnet: B-added Pd$_{0.75}$Mn$_{0.25}$ alloy}

\author{Jiro Kitagawa}
\address{Department of Electrical Engineering, Faculty of Engineering, Fukuoka Institute of Technology, 3-30-1 Wajiro-higashi, Higashi-ku, Fukuoka 811-0295, Japan}
\ead{j-kitagawa@fit.ac.jp}
\author{Kohei Sakaguchi}
\address{Department of Electrical Engineering, Faculty of Engineering, Fukuoka Institute of Technology, 3-30-1 Wajiro-higashi, Higashi-ku, Fukuoka 811-0295, Japan}

\begin{abstract}
Mn-based room-temperature ferromagnets attract considerable attention due to their high ordered Mn moment. We have found that a  Pd$_{0.75}$Mn$_{0.25}$ alloy with added B (Pd$_{0.75}$Mn$_{0.25}$B$_{x}$) shows room-temperature soft ferromagnetism, whereas the parent Pd$_{0.75}$Mn$_{0.25}$ alloy is a spin-glass system. The saturated Mn moment at room temperature systematically increases to 2.68$\mu_{B}$/Mn as $x$ increases to 0.125. The maximum Curie temperature of 390 K is also realized at an $x$ of 0.125. The experimental results suggest a tunable soft ferromagnetism, which is governed only by the boron concentration. Our results will pave the way in chemical control of room-temperature bulk ferromagnetism in Mn compounds based on the addition of an atom with a small atomic radius.
\end{abstract}


\end{frontmatter}

\clearpage

\section{Introduction}
Mn-based ferromagnets attract considerable attention because the Mn atom exhibits the highest ordered moment value\cite{Chikazumi:book,Eriksson:JMMM2007,Francisco:JACS2010,Lamichhane:APL2016}.
Room-temperature Mn-based ferromagnets are particularly important for practical applications.
MnBi is a well-studied room-temperature ferromagnet, showing a high saturation magnetic moment and high magnetic anisotropy\cite{Guillaud:JPR1951,Heikes:PR1955,Yang:APL2001}.
MnBi was considered as a candidate material for magneto-optical memory devices\cite{Williams:1957,Chen:JAP1964}.
MnAl has recently received considerable attention because it shows a rather high magnetic moment and a large magnetic anisotropy energy\cite{Park:JAP2010}.

The Bethe-Slater curve is a useful criterion for understanding the magnetism of 3$d$ transition metal elements\cite{Slater:PR1930-35,Slater:PR1930-36,Sommerfeld:book,Cardias:SR2017}. 
This curve describes a relation between the exchange coupling and the interatomic distance.
For Mn-based compounds, it is generally accepted that a ferromagnetic coupling tends to occur with increasing Mn-Mn distance\cite{Zhu:JAP2005,Simsek:IEEE2015}.
The Bethe-Slater curve motivated us to conduct materials research on Mn-based compounds, offering a platform enabling control of the Mn-Mn distance. 
Regarding control of the Mn-Mn distance, the Mn$_{5}$Si$_{3}$ thin film with added C is interesting.
Although the parent compound Mn$_{5}$Si$_{3}$ is an antiferromagnet with a N\'{e}el temperature of 98 K, Mn$_{5}$Si$_{3}$C$_{x}$ sputtered on a substrate is a room-temperature ferromagnet\cite{Gajdzik:JAP2000,Surgers:PRB2003}.
The unit cell volume expands with increasing carbon content\cite{Surgers:PRB2003}, which means an increase in the Mn-Mn distance.
The ordered Mn moment rapidly increases to 1 $\mu_{B}$/Mn as the carbon content is increased to 0.75.
The appearance of ferromagnetism is attributed to the increased interaction between Mn atoms mediated by the added carbon\cite{Gajdzik:JAP2000}.

The Pd-Mn alloys (Pd$_{1-y}$Mn$_{y}$ with 0$\leq$$y$$\leq$0.25) crystallize into the cubic $\alpha$-Pd structure with space group Fm\={3}m, in which Pd and Mn atoms randomly occupy the Wyckoff 4$a$ site.
The disorder of Pd and Mn atoms leads to a spin-glass behavior\cite{Saha:PRB1994}.
In particular, for $y=$0.25, the heat treatment during sample preparation severely affects the degree of atomic disorder.
A sample annealed at a temperature higher than approximately 500 $^{\circ}$C generally possesses atomic disorder with an $\alpha$-Pd-type structure\cite{Saha:PRB1994}.
The disordered Pd$_{0.75}$Mn$_{0.25}$ exhibits a spin-glass transition\cite{Rashid:JAP1984} at approximately 45 K. 
However, a prolonged annealing below 500 $^{\circ}$C obtains the ordered AuCu$_{3}$-type (or Al$_{3}$Zr-type) Pd$_{3}$Mn, showing an antiferromagnetic ordering at 170$\sim$195 K\cite{Cable:PR1962,Yasui:JDP1988}.
Pd$_{0.75}$Mn$_{0.25}$ alloy with the $\alpha$-Pd-type structure can interstitially incorporate boron atoms\cite{Sakamoto:JAC1993}.
In an earlier report\cite{Sakamoto:JAC1993} of Pd$_{0.75}$Mn$_{0.25}$B$_{x}$ with 0$\leq$$x$$\leq$0.05, although only the $\alpha$-Pd structure was maintained in the $x$ range of 0$\leq$$x$$<$0.0375, a coexistence of the $\alpha$-Pd-type structure and the ordered AuCu$_{3}$-type one was observed in 0.0375$\leq$$x$$\leq$0.05.
The lattice parameters of both structures systematically increase with increasing $x$.

The situation between Mn$_{5}$Si$_{3}$ thin film with added C and Pd$_{0.75}$Mn$_{0.25}$ alloy with added B is similar: the Mn-Mn distance increases with increasing carbon or boron content. This behavior motivated us to investigate the magnetic properties of  Pd$_{0.75}$Mn$_{0.25}$B$_{x}$.
In this paper, we report the sample preparation and characterization of Pd$_{0.75}$Mn$_{0.25}$B$_{x}$ with nominal $x$ ranging from 0 to 0.25 and their magnetic properties.

\section{Materials and methods}
Polycrystalline samples of Pd$_{0.75}$Mn$_{0.25}$B$_{x}$ ($x$=0, 0.025, 0.05, 0.075, 0.1, 0.125, 0.15, 0.175, 0.25) were prepared using Pd powder (99.9\%), Mn powder (99.9\%) and B powder (99\%).
Stoichiometric amounts of the powders were homogeneously mixed and pressed into a pellet, which was arc melted under an Ar atmosphere.
The as-cast sample was placed on a Ta sheet and annealed in an evacuated quartz tube at 800 $^{\circ}$C for 4 days.
The samples were evaluated using a powder X-ray diffractometer (Shimadzu, XRD-7000L) with Cu-K$\alpha$ radiation. 
Due to the rather high ductility of the prepared samples, except for the one with $x=$0.25, we used thin slabs for X-ray diffraction (XRD) measurements.
The atomic compositions of the samples were verified by using an energy-dispersive X-ray (EDX) spectrometer that was equipped in a field-emission scanning electron microscope (FE-SEM; JEOL, JSM-7100F). 
After explaining the EDX analysis results, the nominal $x$ value is replaced by the actual value.

The isothermal magnetization $M$ curve at room temperature was measured by a vibrating sample magnetometer (VSM; Tamagawa Seisakusyo, TM-VSM2330HGC).
The temperature dependence of magnetization under an applied external field $H$ of 10 kOe between 300 K and 450 K (for a few samples between 80 K and 450 K) was checked by another VSM (Riken Denshi, BHV-50H).
The temperature dependence of ac magnetic susceptibility $\chi_{ac}$ (T) in an alternating field of 5 Oe at 800 Hz between 4 K and 300 K was measured using a closed-cycle He gas cryostat.
The temperature dependence of electrical resistivity $\rho$ (T) between 4 and 300 K was measured by the conventional DC four-probe method using the cryostat.

\section{Results and discussion}

\begin{figure}
\begin{center}
\includegraphics[width=10cm]{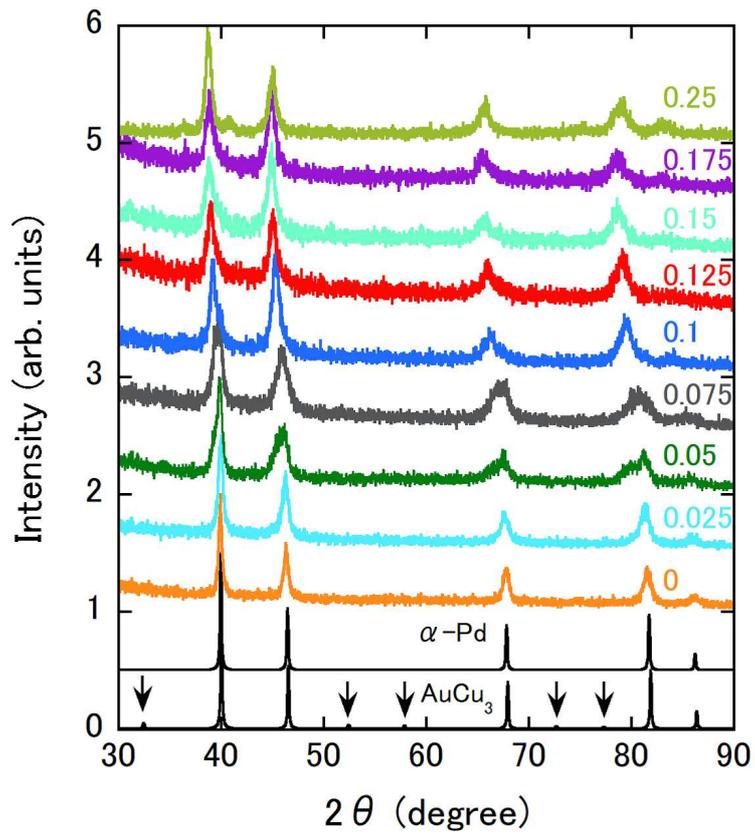}
\end{center}
\caption{XRD patterns of Pd$_{0.75}$Mn$_{0.25}$B$_{x}$. The nominal $x$ values are shown in the figure. The simulated patterns of Pd$_{0.75}$Mn$_{0.25}$ with the $\alpha$-Pd-type and AuCu$_{3}$-type structures are also shown. The origin of each pattern is shifted by 0.5 for clarity.}
\label{f1}
\end{figure}

Figure 1 shows the XRD patterns of the prepared samples with the simulated patterns of Pd$_{0.75}$Mn$_{0.25}$ with the $\alpha$-Pd-type and AuCu$_{3}$-type structures.
The XRD pattern of the AuCu$_{3}$-type structure, which is the ordered variant of  the $\alpha$-Pd-type structure, shows small superlattice reflections denoted by arrows.
Because the peak intensity of the prepared sample is weak, it is difficult to determine which structure type is realized.
However, $\chi_{ac}$ (T) and $\rho$ (T), explained below, support the $\alpha$-Pd-type structure in our sample with $x$=0.
The sample with $x=$0.05 shows rather broad peaks compared to those of the samples with $x=$0 and 0.025.
This result is due to the coexistence of the $\alpha$-Pd-type and AuCu$_{3}$-type structures, as reported in the literature\cite{Sakamoto:JAC1993}.
This coexistence persists in the sample with $x=$0.075, which also presents broad diffraction peaks.
However, further increases in boron content appear to merge the two structures into a single structure, which is supported by the relatively sharper peaks. 
It cannot be concluded which structure type is dominant in the samples with $x\geq$0.1, and further study is needed.
As the boron content increases from $x$=0 to 0.15, each peak position shifts to a lower angle, indicating a lattice expansion.
When $x$ is further increased from 0.15 to 0.25, there is no large shift in angle position. 
The $x$ dependence of the lattice parameter is discussed below after explaining the analysis of atomic composition.

\begin{figure}
\begin{center}
\includegraphics[width=10cm]{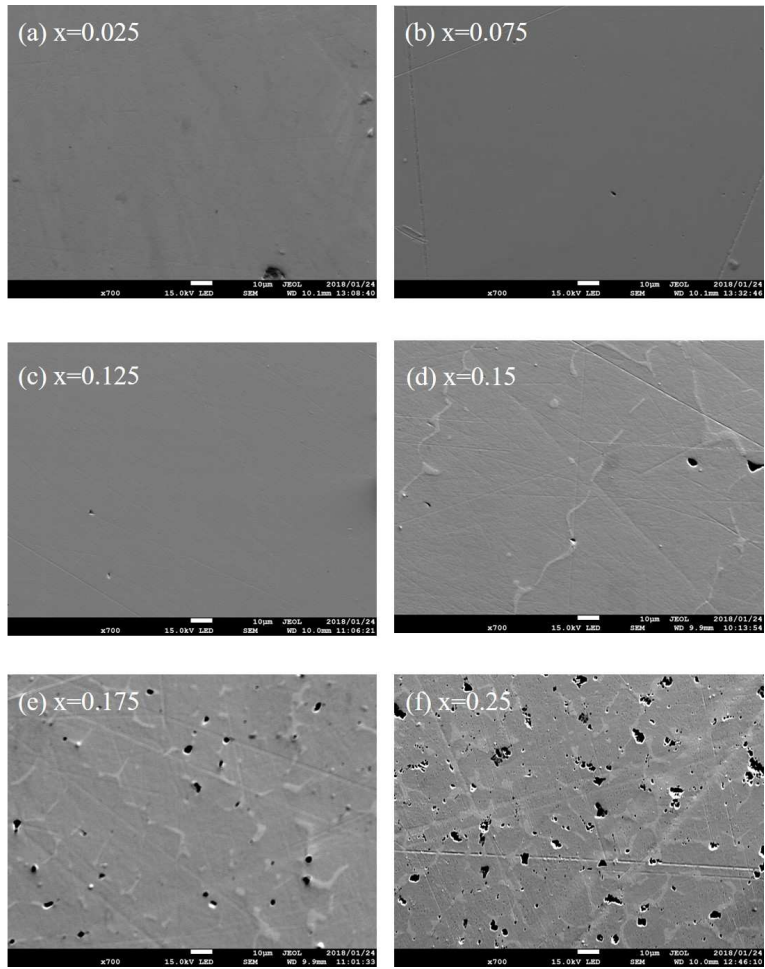}
\end{center}
\caption{Backscattered electron (15 keV) images of Pd$_{0.75}$Mn$_{0.25}$B$_{x}$ for nominal $x$ values of (a) 0.025, (b) 0.075, (c) 0.125, (d) 0.15, (e) 0.175 and (f) 0.25.}
\label{f2}
\end{figure}

\begin{table}[t]
\caption{Nominal atomic composition, atomic composition determined by EDX measurement, lattice parameter, the nearest Mn-Mn distance, saturation moment $M_{s}$ at room temperature (RT) and $T_{C}$ of Pd$_{0.75}$Mn$_{0.25}$B$_{x}$.}
\label{t1}
\begin{tabular}{cccccc}
\hline
\scriptsize{Nominal} & \scriptsize{Determined composition} & \scriptsize{lattice} & \scriptsize{the nearest Mn-Mn} & \scriptsize{$M_{s}$ at RT} & \scriptsize{$T_{C}$} \\
\scriptsize{composition}&& \scriptsize{parameter (\AA)} & \scriptsize{distance (\AA)} & \scriptsize{($\mu_{B}$/Mn)} & \scriptsize{(K)} \\
\hline
\scriptsize{Pd$_{0.75}$Mn$_{0.25}$} & \scriptsize{Pd$_{0.766(1)}$Mn$_{0.233(1)}$} & \scriptsize{3.909(1)} & \scriptsize{2.764} & - & - \\
\scriptsize{Pd$_{0.75}$Mn$_{0.25}$B$_{0.025}$} & \scriptsize{Pd$_{0.763(3)}$Mn$_{0.237(1)}$B$_{0.015(5)}$} & \scriptsize{3.916(2)} & \scriptsize{2.769} & \scriptsize{0.07} & \scriptsize{325(10)} \\
\scriptsize{Pd$_{0.75}$Mn$_{0.25}$B$_{0.05}$} & \scriptsize{Pd$_{0.763(1)}$Mn$_{0.237(1)}$B$_{0.055(3)}$} & \scriptsize{3.925(2)} & \scriptsize{2.775} & \scriptsize{0.87} & \scriptsize{340(5)} \\
\scriptsize{Pd$_{0.75}$Mn$_{0.25}$B$_{0.075}$} & \scriptsize{Pd$_{0.761(3)}$Mn$_{0.239(1)}$B$_{0.070(5)}$} & \scriptsize{3.936(2)} & \scriptsize{2.783} & \scriptsize{1.53} & \scriptsize{339} \\
\scriptsize{Pd$_{0.75}$Mn$_{0.25}$B$_{0.1}$} & \scriptsize{Pd$_{0.765(4)}$Mn$_{0.235(3)}$B$_{0.105(5)}$} & \scriptsize{3.988(2)} & \scriptsize{2.820} & \scriptsize{2.50} & \scriptsize{374} \\
\scriptsize{Pd$_{0.75}$Mn$_{0.25}$B$_{0.125}$} & \scriptsize{Pd$_{0.764(4)}$Mn$_{0.236(3)}$B$_{0.125(5)}$} & \scriptsize{4.008(2)} & \scriptsize{2.834} & \scriptsize{2.68} & \scriptsize{390} \\
\scriptsize{Pd$_{0.75}$Mn$_{0.25}$B$_{0.15}$} & \scriptsize{Pd$_{0.757(2)}$Mn$_{0.243(1)}$B$_{0.148(3)}$} & \scriptsize{4.026(2)} & \scriptsize{2.847} & \scriptsize{2.18} & \scriptsize{330} \\
\scriptsize{Pd$_{0.75}$Mn$_{0.25}$B$_{0.175}$} & \scriptsize{Pd$_{0.761(2)}$Mn$_{0.239(1)}$B$_{0.155(5)}$} & \scriptsize{4.031(2)} & \scriptsize{2.850} & -  & \scriptsize{252} \\
\scriptsize{Pd$_{0.75}$Mn$_{0.25}$B$_{0.25}$} & \scriptsize{Pd$_{0.765(1)}$Mn$_{0.235(3)}$B$_{0.168(5)}$} & \scriptsize{4.020(2)} & \scriptsize{2.843} & -  & \scriptsize{256} \\
\hline
\end{tabular}
\end{table}

\begin{figure}
\begin{center}
\includegraphics[width=8.5cm]{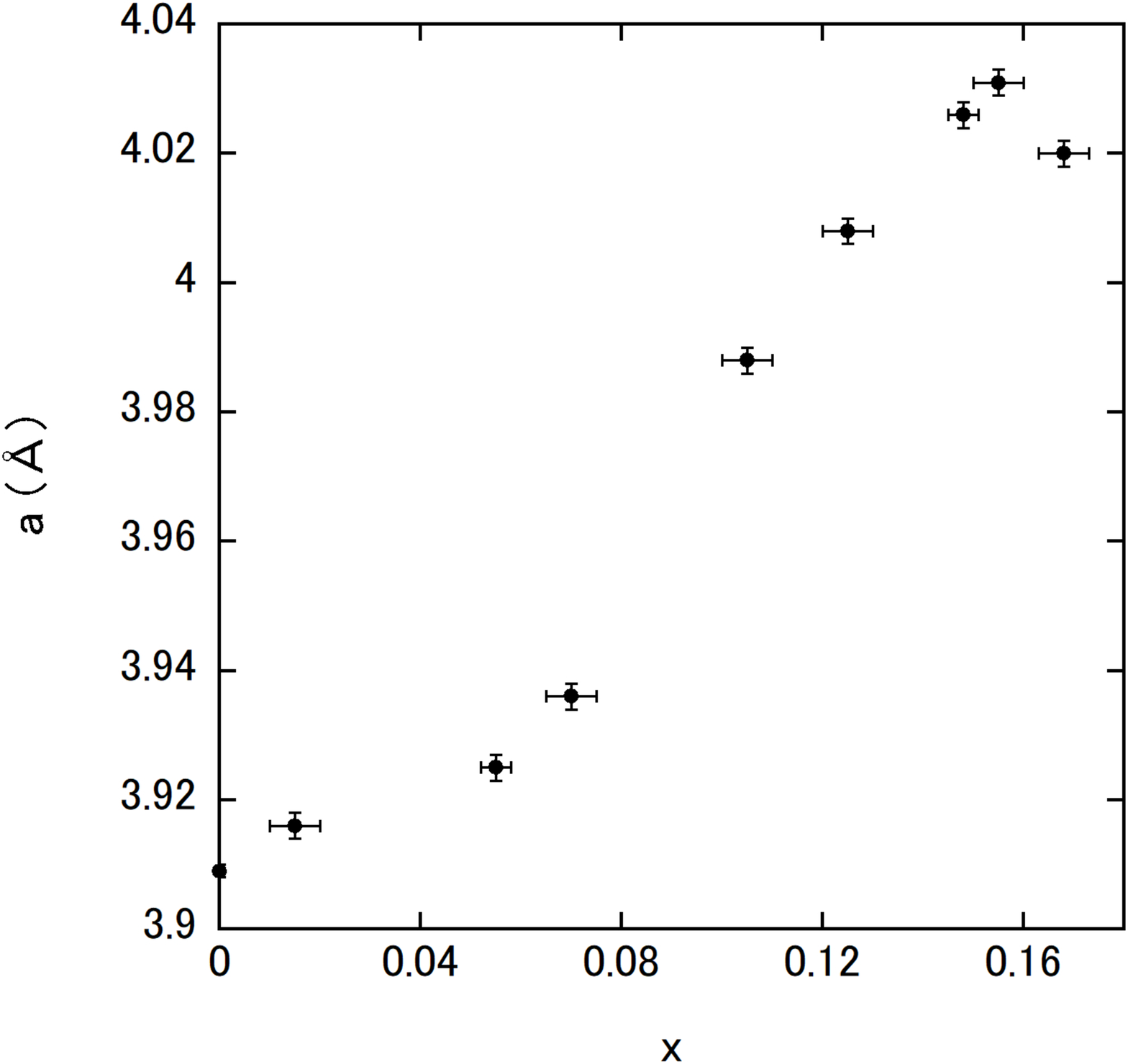}
\end{center}
\caption{Boron content dependence of lattice parameter in Pd$_{0.75}$Mn$_{0.25}$B$_{x}$.}
\label{f3}
\end{figure}

Several backscattered electron images obtained by FE-SEM with electron beams of 15 keV are shown in Fig.\ 2.
For each sample with $x\leq$ 0.125, a noncontrast image means almost a single phase. 
Meanwhile, in the sample with $x$=0.15, a small amount of a brighter image appears, indicating the existence of a secondary phase, the area of which systematically increases with increasing $x$.
This result suggests a B concentration limit in Pd$_{0.75}$Mn$_{0.25}$ alloy.
The atomic composition of the secondary phase is determined to be Pd$_{0.74(1)}$Mn$_{0.09(3)}$B$_{0.19(5)}$.
In the sample with $x$=0.15, voids with small black areas, probably formed during the arc melting, appear.
The void area increases as $x$ increases from 0.15, suggesting some relation between the void formation and the appearance of the secondary phase.
The atomic compositions obtained from EDX measurements of each sample are listed in Table 1.
The respective sample shows the composition ratio between Pd and Mn atoms, which is near the ideal ratio.
The determined B content of 0.015(5) for the nominal composition Pd$_{0.75}$Mn$_{0.25}$B$_{0.025}$ deviates rather greatly from 0.025, probably due to the difficulty of accurately weighing such a small amount of B powder.
In samples with a nominal $x$ between 0.05 and 0.15, the actual boron contents are in good agreement with the nominal ones.
The large difference between the actual and nominal $x$ values is confirmed for the samples with nominal $x$ values of 0.175 and 0.25.
The smaller $x$ obtained by EDX measurement means an incomplete B addition, which is consistent with the FE-SEM images shown in Fig.\ 2.
Hereafter, we use the $x$ values determined by the EDX measurement.

The lattice parameters of the prepared samples were refined by the least squares method using XRD data and are presented in Fig.\ 3 (see also Table 1).
The $\alpha$-Pd-type structure is employed as the crystal structure for all samples.
In each Bragg reflection peak of the samples with $x=$ 0.055 and 0.070, possibly showing the coexistence of $\alpha$-Pd-type and AuCu$_{3}$-type structures, the reflection at higher angles is assumed to be due to the $\alpha$-Pd-type structure, following the assignment in the literature\cite{Sakamoto:JAC1993}.
The lattice parameter $a$ gradually increases as $x$ is increased from 0 to 0.070.
The slope of the $a$ vs $x$ curve becomes steeper between $x=$ 0.070 and 0.148.
Above $x=$ 0.148, $a$ weakly depends on $x$, which might reflect the difficulty of addition with high B concentration mentioned above.
Figure 3 indicates an increase in Mn-Mn distance as $x$ increases from 0 to 0.155.

\begin{figure}
\begin{center}
\includegraphics[width=12cm]{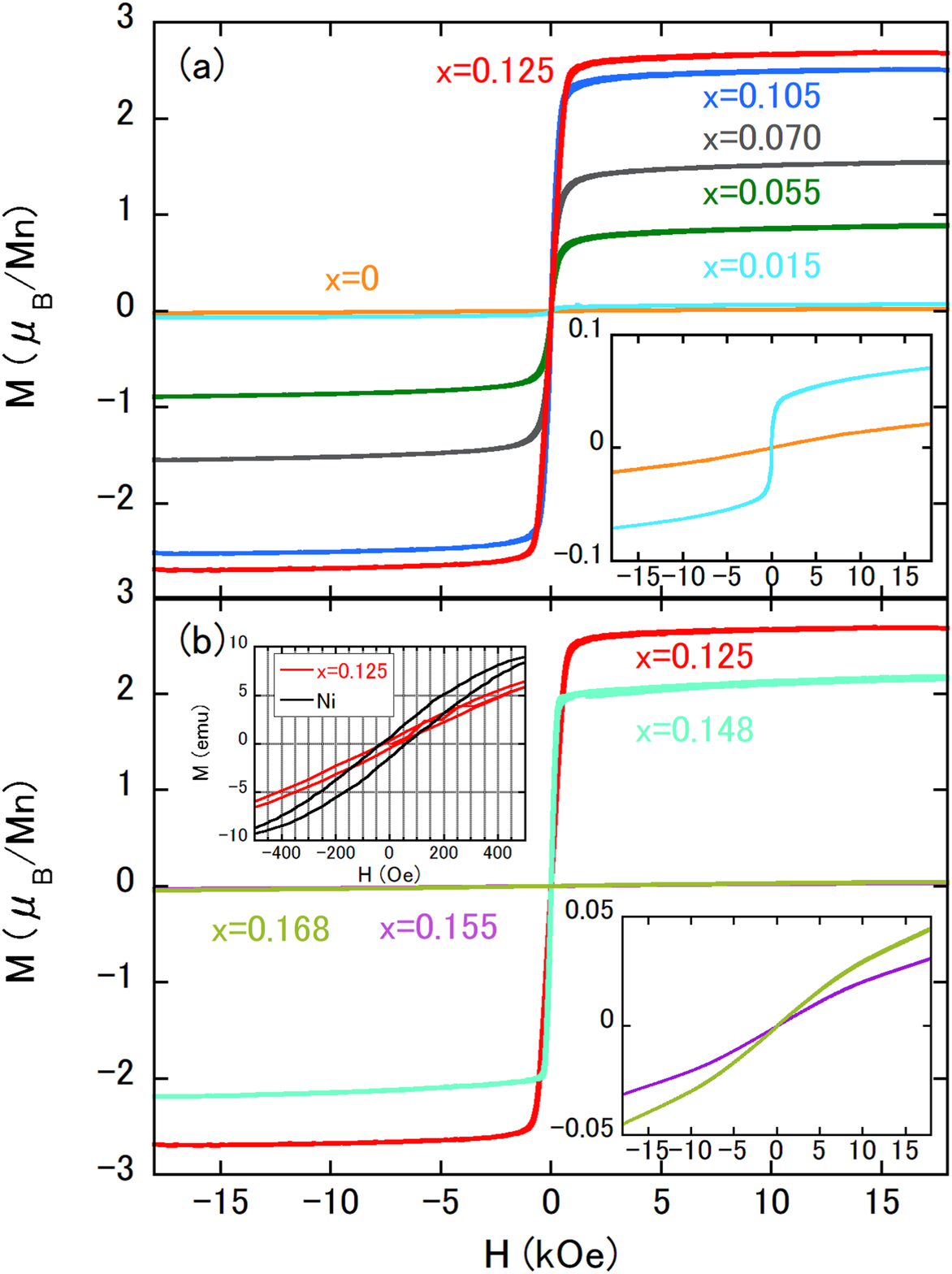}
\end{center}
\caption{Isothermal magnetization curves of Pd$_{0.75}$Mn$_{0.25}$B$_{x}$ at room temperature for (a) $x$=0, 0.015, 0.055, 0.070, 0.105 and 0.125 and for (b) $x$=0.125, 0.148, 0.155 and 0.168. The lower insets in the figures are $M$-$H$ curves with expanded vertical scales of samples with $x$=0 and 0.015 for (a) and $x$=0.155 and 0.168 for (b), respectively. The upper inset in (b) is $M$-$H$ curves with an expanded horizontal scale of the sample with $x$=0.125 and Ni.}
\label{f4}
\end{figure}

Figures 4(a) and 4(b) show the isothermal magnetization curves of Pd$_{0.75}$Mn$_{0.25}$B$_{x}$ at room temperature ($\sim$300 K). 
At $x=$0, 0.155 and 0.168, $M$ is very small and varies almost linearly as a function of $H$, indicating the paramagnetic state (see the lower insets in Figs.\ 4(a) and 4(b)).
The other samples show a weak hysteresis and $M$ saturating at approximately 1 kOe, which are characteristic of a soft ferromagnet, although the saturation of $M$ in the sample with $x$=0.015 appears to be incomplete.
In the upper inset in Fig.\ 4(b), the comparison of coercive fields between the sample with $x$=0.125 and Ni standard used in our VSM measurements is presented. 
The coercive field of the Mn compound is smaller than 50 Oe, and a more precise measurement is needed for discussing a low coercive field.
The saturation magnetization $M_{s}$ systematically increases with increasing $x$ (see Fig.\ 4(a)) and reaches 2.68$\mu_{B}$/Mn for the sample with $x$=0.125, above which $M_{s}$ decreases, as shown in Fig.\ 4(b). 

\begin{figure}
\begin{center}
\includegraphics[width=13cm]{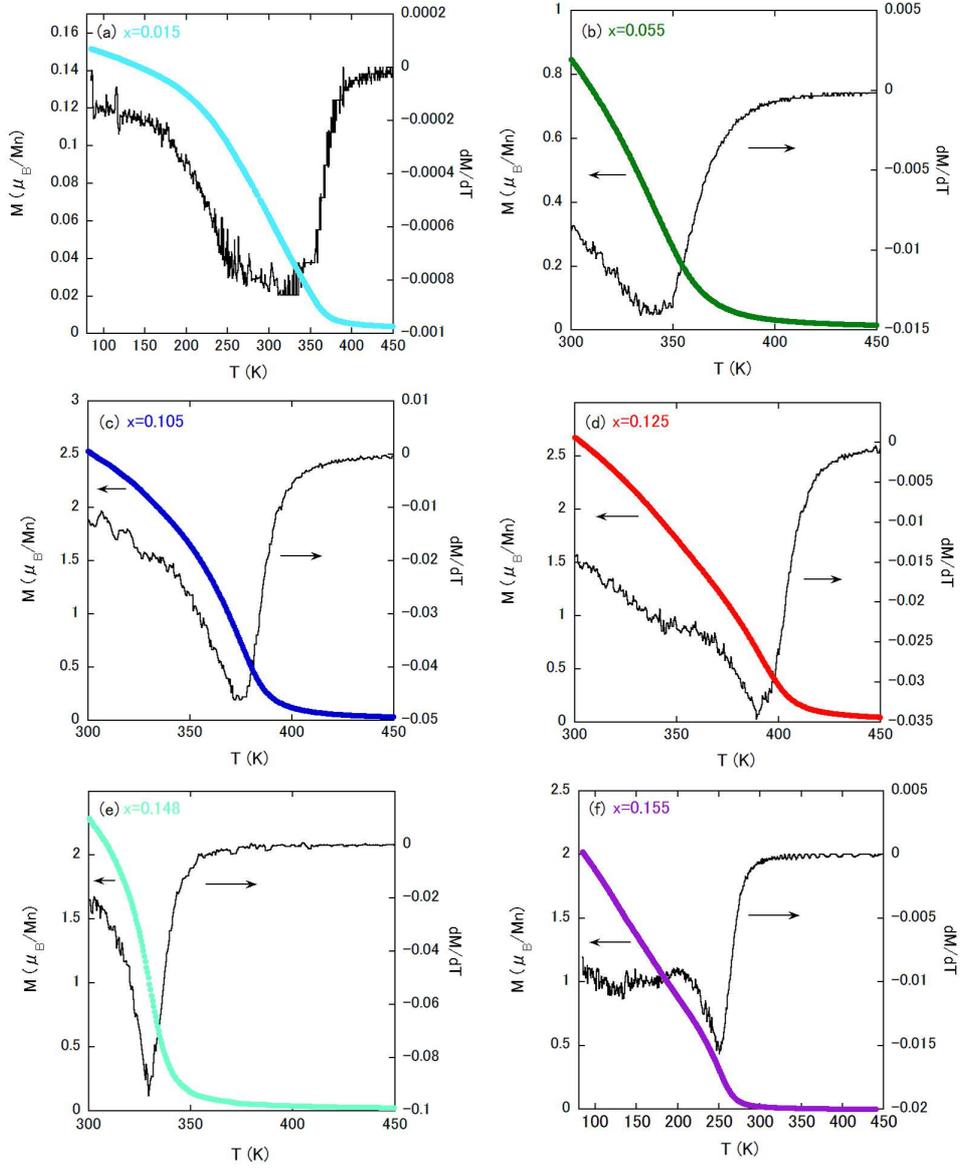}
\end{center}
\caption{Temperature dependences of $M$ measured under an external field of 10 kOe for samples with (a) $x$=0.015, (b) $x$=0.055, (c) $x$=0.105, (d) $x$=0.125, (e) $x$=0.148 and (f) $x$=0.155. $dM/dT$ is also shown in each figure.}
\label{f5}
\end{figure}

The Curie temperature $T_{C}$ was determined by the minimum of the temperature derivative of $M$ under $H$=10 kOe $dM/dT$, as shown in Figs.\ 5(a) to 5(f) for several samples.
This procedure is frequently employed in ferromagnets containing 3$d$ transition metals\cite{Oikawa:APL2001,Yu:APL2003}.
Although a weak increase in $T_{C}$ as $x$ increases from 0.015 to 0.070 is observed (see also Table 1), $T_{C}$ begins to increase rather greatly as $x$ increases from 0.070 to 0.125.
The maximum $T_{C}$ of 390 K is reached at $x$=0.125.
Further increasing $x$ to greater than 0.125 suppresses $T_{C}$.
For a ferromagnetic system, $H$=10 kOe used for the determination of $T_{C}$ might be relatively higher; thus, we have checked $T_{C}$ by measuring other physical quantities $\chi_{ac}$ (T) and $\rho$ (T) for the sample with $x$=0.155.
As shown in Fig.\ 6(a), $\chi_{ac}$ (T) increases steeply below approximately $T_{C}$. 
Figure 6(b) displays the comparison of $d\chi_{ac}/dT$ and $dM/dT$, which supports the validity of determining $T_{C}$ by magnetization measurement under $H$=10 kOe for the present system.
Further support is provided by the metallic $\rho$, showing a kink at $T_{C}$, where $dM/dT$ also reaches a minimum value (see Fig.\ 6(b)).

\begin{figure}
\begin{center}
\includegraphics[width=11cm]{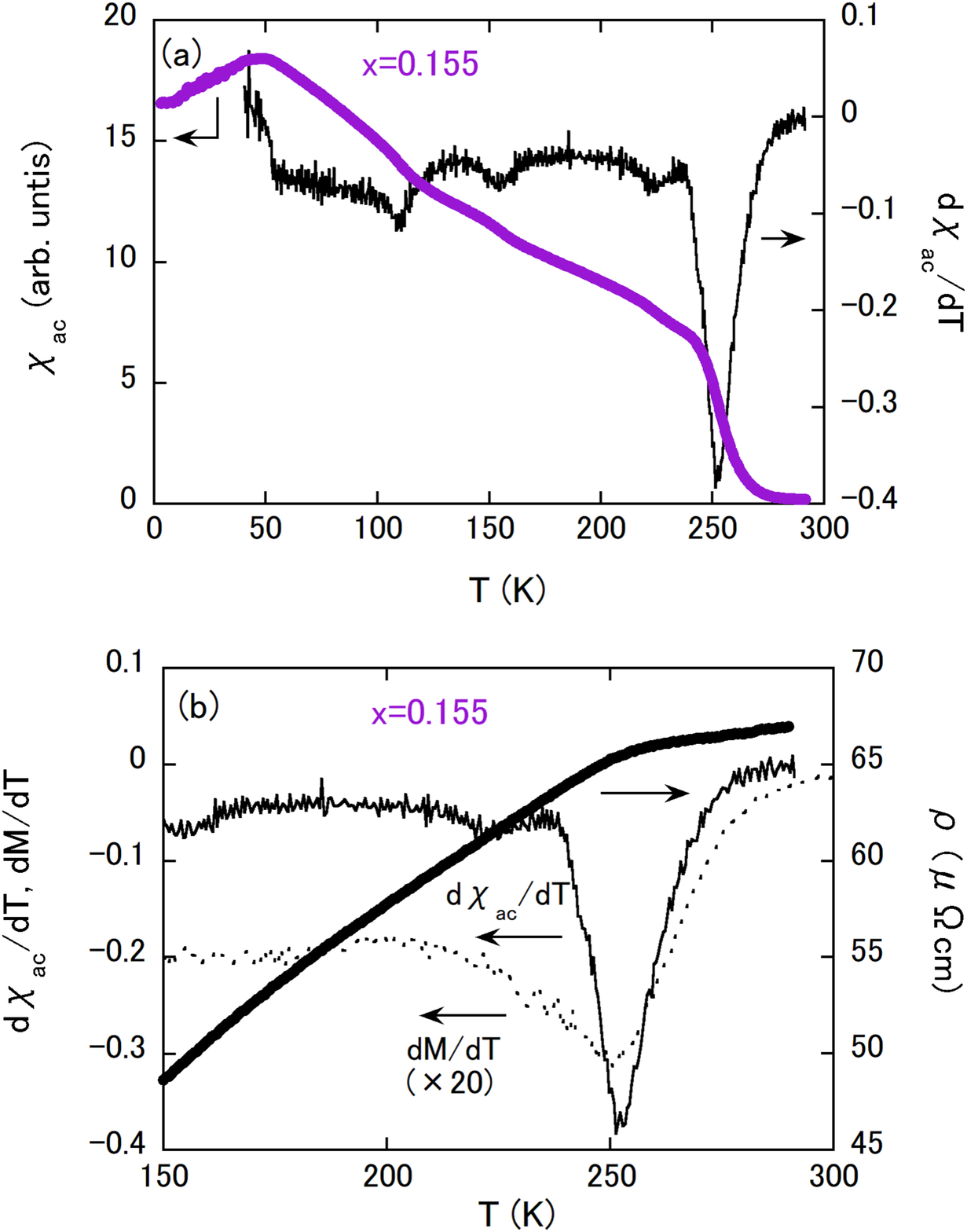}
\end{center}
\caption{(a) Temperature dependences of $\chi_{ac}$ and $d\chi_{ac}/dT$ of the sample with $x$=0.155. $d\chi_{ac}/dT$ below approximately 40 K is eliminated because it is very noisy. (b) Comparison of $d\chi_{ac}/dT$ and $dM/dT$ and temperature dependence of $\rho$ for the sample with $x$=0.155.}
\label{f6}
\end{figure}

More detailed magnetic properties are investigated by $\chi_{ac}$ (T) measurements for all samples, as shown in Figs.\ 7(a) to 7(e). 
Fig.\ 7(a) is the $\chi_{ac}$ (T) of the parent sample ($x$=0), showing a cusp at approximately 45 K, which is the same as the reported magnetic susceptibility of disordered Pd$_{0.75}$Mn$_{0.25}$ interpreted as a spin-glass system\cite{Rashid:JAP1984}.
As shown in the inset of Fig.\ 7(a), $\rho$ (T) of the parent sample exhibits a very weak temperature dependence, which is also consistent with the disordered $\alpha$-Pd-type compound.
From room temperature to approximately 50 K, the $\chi_{ac}$ (T) of each ferromagnetic sample continuously increases, except for the small humps at 230$\sim$260 K in the samples with $x$=0.015, 0.055 and 0.070.
$\chi_{ac}$ (T) is correlated with the reversible initial magnetization process.
The continuous increase in $\chi_{ac}$ (T) below $T_{C}$ suggests an easily reversible magnetization process, accommodating the nature of soft ferromagnets.
$\chi_{ac}$ (T) of the sample with $x$=0.015 shows a small hump, signaling a magnetic phase transition at 230 K, which shifts to 260 K in the sample with $x$=0.055.
However, the hump is suppressed in the sample with $x$=0.070.
The XRD patterns of the samples with $x$=0.055 and 0.070 suggest the coexistence of the $\alpha$-Pd-type and AuCu$_{3}$-type structures.
Therefore, either of the structures would be responsible for the room temperature ferromagnetism and the other one for the small hump.
$\chi_{ac}$ (T) has revealed that the coexistence of two types of structures is also realized in the sample with $x$=0.015, which is not clearly detected by the XRD pattern.

\begin{figure}
\begin{center}
\includegraphics[width=13cm]{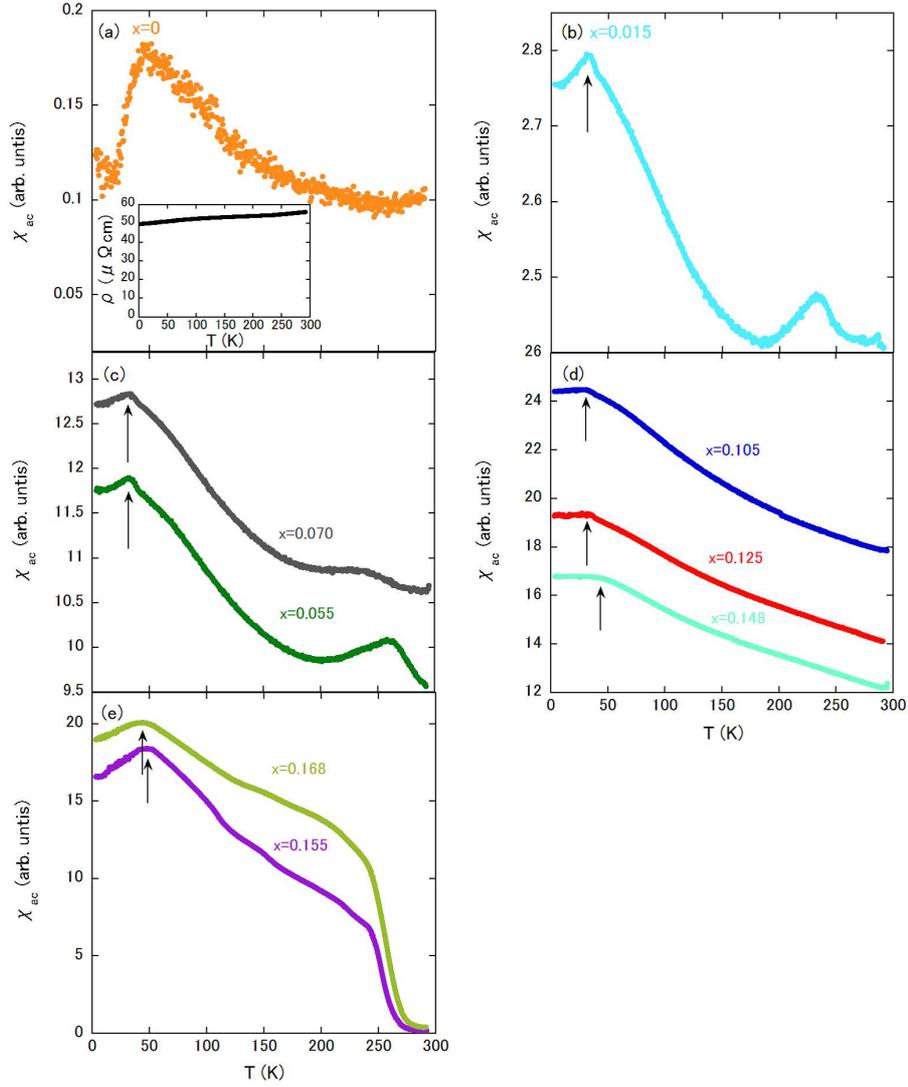}
\end{center}
\caption{Temperature dependences of $\chi_{ac}$ for samples with (a) $x$=0; (b) $x$=0.015; (c) $x$=0.055 and 0.070; (d) $x$=0.105, 0.125 and 0.148; and (e) $x$=0.155 and 0.168. The inset of (a) is $\rho$ (T) of the parent sample ($x$=0).}
\label{f7}
\end{figure}

The ferromagnetic samples of Pd$_{0.75}$Mn$_{0.25}$B$_{x}$ show peaks or shoulders below 50 K (see the arrows in Figs.\ 7(b) to 7(e)), which is close to the spin-glass transition temperature in the parent sample.
The origin of peaks or shoulders is an open question, and further detailed studies are needed; however, we comment on this issue.
In some spin-glass systems, breaking the competition between antiferromagnetic and ferromagnetic exchange interactions toward a dominance of the ferromagnetic one by changing the concentration of magnetic species induces a ferromagnetic state.
The incipient stage of the development of ferromagnetism frequently accompanies a spin-glass state below $T_{C}$, which is called a reentrant spin glass.
In Pd$_{0.75}$Mn$_{0.25}$B$_{x}$, the results of $\chi_{ac}$ (T) evoke an occurrence of a reentrant spin glass.
Indeed, as $x$ is increased from 0 to 0.055, accompanying the appearance of ferromagnetism and the slight enhancement of $T_{C}$, the temperature showing the $\chi_{ac}$ peak shifts to the lower-temperature side, which is consistent with typical magnetic phase diagrams of reentrant spin-glass systems\cite{Yeshurun:PRL1980,Mamchik:PRB2004}.
However, the further increased $T_{C}$ does not severely affect the peak (shoulder) position (see Fig.\ 7(d)), which might be inconsistent with the reentrant spin glass. 
Note that for the samples with $x\geq$0.148, the temperatures at peaks (shoulder) are higher than that of the sample with $x$=0.125.
This result may indicate that the reduction of $T_{C}$ at $x\geq$0.148 is correlated with the enhancement of the peak showing temperature (shoulder).
In the present stage, it cannot be excluded that the spin-glass state in the parent compound has a relationship with the peak (shoulder) of $\chi_{ac}$ (T) in another sample.

$M_{s}$ and $T_{C}$ of each sample are summarized in Table 1 and plotted as a function of $x$ in Figs.\ 8(a) and 8(b), respectively.
In the samples with $x\leq$ 0.148, $M_{s}$ is positively correlated with $T_{C}$. 
The comparison of ferromagnetic properties between Pd$_{0.75}$Mn$_{0.25}$B$_{x}$ and Mn$_{5}$Si$_{3}$ thin film with added carbon is discussed below.
The maximum $M_{s}$ of 2.68 $\mu_{B}$/Mn in Pd$_{0.75}$Mn$_{0.25}$B$_{0.125}$ at room temperature already exceeds that of 1 $\mu_{B}$/Mn in Mn$_{5}$Si$_{3}$C$_{0.75}$ obtained at 5 K.
The difference of $M_{s}$ between the two systems would be responsible for that of the Mn moment value between the parent compounds Pd$_{0.75}$Mn$_{0.25}$ and Mn$_{5}$Si$_{3}$.
The antiferromagnetic structure of Mn$_{5}$Si$_{3}$ has been clarified by a neutron diffraction study\cite{Brown:JPCM1992}.
There are two inequivalent crystallographic  sites for Mn atoms: the 4$d$ and 6$g$ sites.
The averaged Mn moment value of the two sites is calculated to be 1.31 $\mu_{B}$/Mn.
Although the Mn moment value of disordered Pd$_{0.75}$Mn$_{0.25}$ is unknown, the magnetic structural study\cite{Cable:PR1962} of the ordered AuCu$_{3}$-type Pd$_{3}$Mn has reported a Mn moment value of 4.0 $\mu_{B}$/Mn.
Therefore, it can be speculated that the Mn moment value of the parent compound would be reflected in that of the B(C)-added samples.
The composition dependences of $T_{C}$ for both systems are somewhat different.
Mn$_{5}$Si$_{3}$C$_{x}$ thin film has a wide composition range, showing a $T_{C}$ plateau at the maximum $T_{C}$ of approximately 350 K.
Conversely, the maximum $T_{C}$ would be realized in a narrow composition range for Pd$_{0.75}$Mn$_{0.25}$B$_{x}$.
$T_{C}$ is a rough indication of the strength of exchange coupling between Mn atoms.
The weak composition dependence of $T_{C}$ in Mn$_{5}$Si$_{3}$C$_{x}$ might suggest a small influence of carbon atoms on the enhanced exchange coupling.
In Pd$_{0.75}$Mn$_{0.25}$B$_{x}$, boron atoms would sensitively affect the strength of exchange coupling between Mn atoms.

\begin{figure}
\begin{center}
\includegraphics[width=12cm]{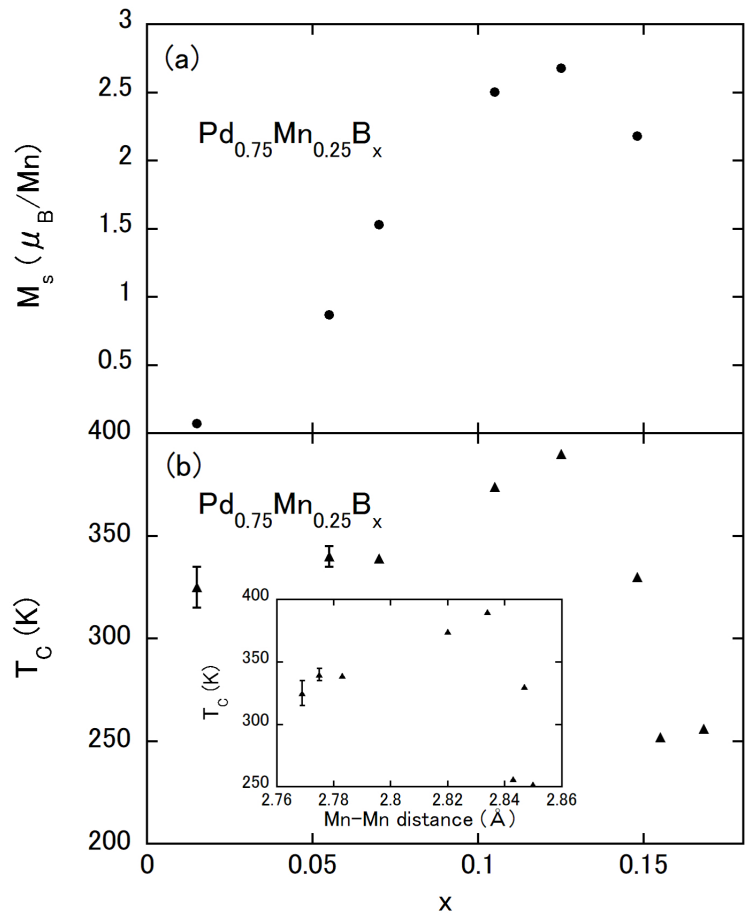}
\end{center}
\caption{Boron content dependence of (a) $M_{s}$ at room temperature and (b) $T_{C}$ of Pd$_{0.75}$Mn$_{0.25}$B$_{x}$. The inset of (b) is $T_{C}$ versus the nearest Mn-Mn distance plot.}
\label{f8}
\end{figure}

The origin for the appearance of ferromagnetism in Pd$_{0.75}$Mn$_{0.25}$B$_{x}$ is unclear at the present stage.
However, some speculations are mentioned below.
The Bethe-Slater curve tells us that an increased Mn-Mn distance favors a ferromagnetic exchange between Mn atoms.
In the inset in Fig.\ 8(b), $T_{C}$ is plotted as a function of the nearest Mn-Mn distance, calculated by assuming the disordered $\alpha$-Pd-type structure (see also Table 1).
The rather good linearity between $T_{C}$ and the Mn-Mn distance ($\leq$ 2.84 \AA) indicates that the B addition with increased Mn-Mn distance enhances the ferromagnetic exchange coupling according to the concept of the Bethe-Slater curve.
Meanwhile, the further increase in Mn-Mn distance above 2.84 \AA \hspace{1pt} reduces $T_{C}$, which means a weakened ferromagnetic interaction.
Therefore, the degree of ferromagnetic exchange coupling would not be determined only by the Mn-Mn distance, and additional factors should be considered.

Boron atoms generally act as donors in Mn compounds\cite{Chikazumi:book}, and a change in electron number is expected.
If the $x$ dependence of $M_{s}$ can be regarded as the electron-number dependence of $M_{s}$, then the result of Fig.\ 8(a) resembles a so-called Slater-Pauling curve\cite{Chikazumi:book,Dederichs:JMMM1991}, which is highly useful for designing ferromagnetic materials based on 3$d$ transition metals.
Because the Pd$_{0.75}$Mn$_{0.25}$B$_{x}$ system shows a rather low $T_{C}$ and contains the precious metal Pd, the system would not be immediately useful for practical applications.
However, our results pave the way in chemical control of room-temperature bulk ferromagnetism in Mn compounds based on the addition of an atom with a small atomic radius.

\section{Summary}
We have found that Pd$_{0.75}$Mn$_{0.25}$B$_{x}$ alloys, in which the parent Pd$_{0.75}$Mn$_{0.25}$ is a spin-glass system, show room-temperature soft ferromagnetism. 
A B concentration limit ($x\sim$ 0.16) exists. 
The lattice parameter increases as $x$ increases from 0 to 0.155, which means that the  Mn-Mn distance systematically increased.
Room-temperature ferromagnetism is observed in the $x$ range between 0.015 and 0.148.
The samples with $x\geq$ 0.148 show ferromagnetic transitions below 300 K.
The saturation moment at room temperature systematically increases as $x$ is increased from 0.015 and reaches 2.68 $\mu_{B}$/Mn at $x=$ 0.125.
The maximum $T_{C}$ is also realized at the boron content.
Pd$_{0.75}$Mn$_{0.25}$B$_{x}$ alloys offer a valuable platform where room-temperature bulk ferromagnetism can be tuned by changing only the concentration of an interstitial nonmagnetic element with a small atomic radius.

\section*{Acknowledgments}
This work was supported by the Fukuoka Institute of Technology's Comprehensive Research Organization.

\end{document}